\documentclass[twocolumn,showpacs,preprintnumbers,amsmath,amssymb]{revtex4}
\usepackage{graphicx}

\newcommand{\be}{\begin{equation}}
\newcommand{\ee}{\end{equation}}
\newcommand{\ba}{\begin{eqnarray}}
\newcommand{\ea}{\end{eqnarray}}
\newcommand{\ds}{\displaystyle}
\newcommand{\MPl}{M_{\mathrm{Pl}}}
\newcommand{\MEW}{M_{\mathrm{EW}}}
\newcommand{\hs}{\hspace{-1mm}}
\newcommand{\eq}{\hs & = & \hs}
\newcommand{\eqequiv}{\hs & \equiv & \hs}
\newcommand{\eqsimeq}{\hs & \simeq & \hs}
\newcommand{\vs}{\vspace{1.5mm}}
\newcommand{\Mathematica}{Mathematica}

\begin{document}

\title{Corrections to the Newtonian potential\\
in the two-brane Randall-Sundrum model}

\author{Petter Callin}
  \email{n.p.callin@fys.uio.no}
\affiliation{Department of Physics, University of Oslo, N-0316 Oslo, Norway\\}

\date{December 9, 2004}

\begin{abstract}
We calculate the Newtonian potential in the two-brane Randall-Sundrum model,
emphasizing the effect of the finite distance between the two branes. The
result obtained is quite natural: When the distance in the potential is small
compared to the brane separation the two-brane model is indistinguishable from
the one-brane model, whereas when the distance is large the bulk dimension
behaves like an ordinary compact dimension, with an exponentially decreasing
correction to the four-dimensional potential. The contribution from the radion
is also included, and is found to give only a multiplicative factor of order
$1$ in the correction.
\end{abstract}


\maketitle

\section{Introduction}

In this paper we study the Randall-Sundrum I model \cite{RandallSundrum1},
which consists of two 3-branes embedded in five-dimensional anti de Sitter
space. Our main focus will be the static gravitational potential between two
point particles on the visible brane, and in particular the effect the finite
gap between the two branes has on the potential compared to the Randall-Sundrum
II model \cite{RandallSundrum2} with only a single brane.

The two-brane model is mostly used in connection with particle physics, where
it gives a natural explanation of the hierarchy between the electroweak scale
and the Planck scale, without the need to introduce very large (or very small)
dimensionless parameters in the theory \cite{RandallSundrum1}. For a survey of
the phenomenology of theories with extra dimensions, see e.g.
\cite{Rubakov_Kubyshin} and references therein. As for the problem of graviton
localization and the derivation of the gravitational potential, most of the
recent activity has focused on the one-brane model \cite{Div_en_bran, Callin}.
However, some work has also been done with two branes in this context, see e.g.
\cite{Smolyakov, Boos}, and \cite{Shtanov} where induced curvature terms in the
actions for the two branes are considered. The cosmology of the two-brane model
is studied in e.g. \cite{Brevik}, and a mechanism for stabilizing the brane
separation is discussed in \cite{GoldbergerWise}. In this paper we will simply
assume that the branes are static and fixed at a constant distance.

Before doing any calculations, we should be able to give a qualitative
description of what the potential should look like. The deciding factor must be
the ratio between the distance $r$  in the potential and the separation between
the two branes. For short distances the second brane is far away, relatively
speaking, and should therefore be invisible to the potential. The second brane
could just as well be sent away to infinity, meaning that the potential must be
the same as in the one-brane model for short distances. Of course, if the
distance $r$ is very short we can also ignore the entire five-dimensional
curvature, meaning that $V(r) \sim 1/r^2$ when $r \to 0$. On the other hand,
for large distances the more detailed structure of the fifth dimension must be
invisible to the potential, and we should get the same result as with an
ordinary, compact extra dimension. That is, the potential should essentially be
four-dimensional, $V(r) \sim 1/r$, with an exponentially decreasing correction
from the fifth dimension.

However, it is not the physical distance $y_r$ to the second brane (as measured
with standard rods) that sets this scale, but rather the conformal distance
$z_r \sim e^{\mu y_r}$. This can be understood from the fact that all particle
masses as measured on the visible brane are reduced by a factor $e^{-\mu y_r}$
as compared to the masses on the hidden brane \cite{RandallSundrum1}. The
exponential factor originates from the warped geometry of the fifth dimension.
Since the gravitational force is mediated by massive gravitons when observed in
four dimensions, and massive propagators result in a factor $e^{-mr}$, the
distance scale is determined by $r \sim m^{-1} \sim e^{+\mu y_r} \sim z_r$,
i.e. the conformal rather than the physical brane separation.

This paper is organized as follows: First, the model is presented in section
\ref{sec:model}. In section \ref{sec:gravitonpropagation} we summarize the
governing equations for graviton propagation in this particular background,
using the results of \cite{Callin}. We then proceed to find the spectrum of
graviton masses in section \ref{sec:Newtonslaw}, allowing us to finally derive
approximate expressions for the gravitational potential in the limit of short
and large distances, respectively.

\section{The model}
\label{sec:model}

We assume that the five-dimensional spacetime can be parametrized by the metric
\be
  ds^2 = A^2(y) \eta_{\mu\nu} dx^\mu dx^\nu - dy^2 ,
\ee
where $A(y)$ is some function of the coordinate $y$ of the fifth dimension,
called the warp factor. The signature we use is $\eta_{\mu\nu} = (+1, -1, -1,
-1)$. The two branes are located at $y=0$ and $y=y_r$ with an orbifold symmetry
$S^1 / Z_2$ in the $y$-direction, and we assume that the only source of gravity
is a cosmological constant $\Lambda_B$ in the bulk, and a tension $\lambda$ and
$\lambda_r$ on the two branes. The warp factor $A(y)$ is determined from the
Einstein equation just as in the one-brane case. With the fine-tuning
$\Lambda_B = -M^{-6} \lambda^2 / 6$, $M$ being the five-dimensional Planck
mass, the solution is
\be
  A(y) = e^{-\mu |y|} ,
  \label{eq:warpfactor}
\ee
where $\mu = \sqrt{|\Lambda_B|/6} = M^{-3}\lambda/6$. The bulk space is anti de
Sitter since $\Lambda_B < 0$, and the fine-tuning means that the effective
four-dimensional cosmological constant vanishes. This configuration is often
called a critical brane.

The two branes impose the boundary conditions
\be
  \left.\frac{[A']}{A}\right|_{y=0} \!\!\!\! =
    -\tfrac{1}{3}M^{-3} \lambda \, , \hspace{5mm}
  \left.\frac{[A']}{A}\right|_{y=y_r} \!\!\!\!\! =
    -\tfrac{1}{3}M^{-3} \lambda_r
\ee
on the warp factor, where $[A']_{y=0} = A'(0^+) - A'(0^-) = 2A'(0^+)$ and
$[A']_{y=y_r} = A'(y_r^+) - A'(y_r^-) = -2A'(y_r^-)$ are the jump
discontinuties of $A'(y)$ across the two branes. The boundary condition at
$y=0$ simply fixes the solution (\ref{eq:warpfactor}), whereas the condition at
$y=y_r$ imposes a constraint on the tension $\lambda_r$ of the second brane.
With $A(y) = e^{-\mu|y|}$ for $|y| \leq y_r$ we get $\lambda_r = -6\mu M^3 =
-\lambda$, which is the well-known result that the two branes have tensions
with the same magnitude but opposite sign. The physical brane where we are
supposed to be living, is usually taken to be the one with a positive tension,
i.e. $\lambda
> 0$ and $\lambda_r < 0$.

A preliminary expression for the four-dimensional Planck mass, ignoring the
contribution from the radion, can be found by integrating over the
$y$-dimension in the five-dimensional gravitational action $S =
-\tfrac{1}{2}M^3 \int d^4 x \, dy \sqrt{\hat{g}} \hat{R}$. All five-dimensional
quantities are here denoted with a hat. The five-dimensional curvature scalar
$\hat{R}$ is related to the four-dimensional one through $\hat{R} = (1/A^2) R +
\ldots$ (replacing for a moment the flat four-dimensional metric
$\eta_{\mu\nu}$ by $g_{\mu\nu}$). Using $\sqrt{\hat{g}} = A^4 \sqrt{g}$ and
demanding that $S = -\tfrac{1}{2}\MPl^2 \int d^4 x \sqrt{g} R + \ldots$, we
then get
\be
  \MPl^2 = M^3 \int_{-y_r}^{y_r} A^2(y) dy =
    \frac{M^3}{\mu} \left( 1 - e^{-2\mu y_r} \right) .
  \label{eq:4dPlanckmass}
\ee
The four-dimensional Planck mass thus depends only weakly on $y_r$ as discussed
in \cite{RandallSundrum1}, and the parameters $M$ and $\mu$ can be chosen to be
of the same order of magnitude as $\MPl$. The presence of the radion changes
this expression slightly, see eq. (\ref{eq:4dPlanckmass_radion}).

\section{Graviton propagation}
\label{sec:gravitonpropagation}

We study the propagation of gravitons in the model by looking at a perturbation
\be
  ds^2 = A^2(y) \left( \eta_{\mu\nu} + h_{\mu\nu} \right) dx^\mu dx^\nu - dy^2
\ee
to the metric \cite{radion}. This analysis is exactly the same for two branes
as for a single brane, only that the regulator brane in \cite{Callin} now
remains at a finite distance. We will therefore only summarize the main results
here, referring to \cite{Callin} for the more detailed calculations.

Imposing the usual gauge conditions $\partial^\mu h_{\mu\nu} = 0$ and
$\eta^{\mu\nu} h_{\mu\nu} = 0$, the wave equation for $h_{\mu\nu}$ reduces to
\be
  \left(
    \frac{1}{A^2} \nabla_\alpha^2 - \partial_y^2 - \frac{4A'}{A} \partial_y
  \right) h_{\mu\nu} = 0 \, ,
\ee
where $\nabla_\alpha^2$ is the four-dimensional Laplacian, $\nabla_\alpha^2 =
\partial_\alpha \partial^\alpha$ in flat space. The separation of variables
$h_{\mu\nu}(x,y) = G_{\mu\nu}(x) \Phi(y)$ yields
\ba
  \left( \nabla_\alpha^2 + m^2 \right) G_{\mu\nu}(x) \eq 0 \, , \\
    \label{eq:4dgraviton}
  \Phi''(y) + \frac{4A'}{A} \Phi'(y) + \frac{m^2}{A^2} \Phi(y) \eq 0 \, .
    \label{eq:5dpart}
\ea
The tensor $G_{\mu\nu}(x)$ describes massive spin-2 particles in four
dimensions, and the allowed values of the mass $m$ are determined from
(\ref{eq:5dpart}). Changing to the conformal coordinate $z$ where $\partial y /
\partial z = A(y)$, meaning that $1+\mu|z| = e^{\mu|y|}$, and writing
$\Phi(y) = A^{-3/2} u(z)$, (\ref{eq:5dpart}) is reduced to
\be
  \left[ -\partial_z^2 + V(z) \right] u(z) = m^2 u(z) \, ,
  \label{eq:Scrodinger}
\ee
where
\ba
  V(z) \eq \frac{9}{4}(A')^2 + \frac{3}{2}AA'' \nonumber \\
  \eq \frac{15\mu^2}{4(1+\mu|z|)^2} - 3\mu\delta(z) +
    \frac{3\mu}{1+\mu z_r} \delta(z-z_r) \, . \hspace{7mm}
  \label{eq:Scr_potential}
\ea
($A'$ still means the derivative of $A$ with respect to $y$.) The two delta
functions in the Schr{\"o}dinger potential result in the boundary conditions
\ba
  2u'(0) + 3\mu u(0) \eq 0 \, ,
    \label{eq:boundary0} \\
  2u'(z_r) + \frac{3\mu}{1+\mu z_r} u(z_r) \eq 0 \, .
    \label{eq:boundaryr}
\ea
The solutions to (\ref{eq:Scrodinger}) with the first boundary condition
(\ref{eq:boundary0}) imposed are
\ba
  u_0(z) \eq N_0 (1 + \mu|z|)^{-3/2} , \nonumber \\
  u_m(z) \eq N_m \sqrt{1 + \mu|z|} \nonumber \\
  && \hspace{-4mm} \times \! \left\{
    Y_2[\tfrac{m}{\mu}(1+\mu|z|)] -
    \frac{Y_1(\frac{m}{\mu})}{J_1(\frac{m}{\mu})} J_2[\tfrac{m}{\mu}(1+\mu|z|)]
  \right\}, \nonumber \\
  && \label{eq:u_solution}
\ea
where $N_0$ and $N_m$ are normalization constants, and $J_n(x)$ and $Y_n(x)$
are Bessel functions of order $n$. The second boundary condition
(\ref{eq:boundaryr}) determines the allowed mass eigenvalues, and can be
simplified to
\be
  \frac{Y_1(\tfrac{m}{\mu})}{J_1(\tfrac{m}{\mu})} =
    \frac{Y_1[\tfrac{m}{\mu}(1+\mu z_r)]}{J_1[\tfrac{m}{\mu}(1+\mu z_r)]}
  \label{eq:eigenvalues}
\ee
for the massive modes. The massless mode will always be present.

\section{Corrections to Newton's law}
\label{sec:Newtonslaw}

As shown in \cite{Callin}, the gravitational potential between two point
particles on the physical brane at $y=0$, due to the exchange of virtual
gravitons, can be written
\be
  V(\mathbf{k}) = \left.
    \frac{1}{M^3} \sum_m |u(m,0)|^2
    \frac{T_1^{\mu\nu}P^{(m)}_{\mu\nu\alpha\beta}T_2^{\alpha\beta}}{k^2-m^2}
  \right|_{k^0 \to 0}.
  \label{eq:V_momentum}
\ee
The energy momentum tensors are $T_1^{\mu\nu} = m_1 u^\mu u^\nu = m_1
\delta^\mu_0 \delta^\nu_0$ and $T_2^{\alpha\beta} = m_2 \delta^\alpha_0
\delta^\beta_0$, and $P^{(m)}_{\mu\nu\alpha\beta}$ is the polarization tensor
of a spin-2 particle in four dimensions with mass $m$ and momentum $k$. The
normalization constants in (\ref{eq:u_solution}) are found by requiring
$\int_{-z_r}^{z_r} u_i(z) u_j(z) dz = \delta_{ij}$, which means that
\be
  N_0^2 = \left[ \int_{-z_r}^{z_r} \frac{dz}{(1+\mu|z|)^3} \right]^{-1} =
    \frac{\mu}{1-e^{-2\mu y_r}} \, .
\ee
Using $P^{(0)}_{0000} = \frac{1}{2}$ and $P^{(m>0)}_{0000} = \frac{2}{3}$,
together with $\MPl^{-2} = 8\pi G$ where $G$ is the Newtonian gravitational
constant, we get after Fourier inverting (\ref{eq:V_momentum})
\ba
  V(r) \eq -\frac{G m_1 m_2}{r} \nonumber \\
  && \hspace{-4mm} \times \! \left\{
    1 + \frac{4(1-e^{-2\mu y_r})}{3\mu} \sum_{m>0} |u(m,0)|^2 e^{-mr}
  \right\}. \hspace{7mm}
  \label{eq:V_eksakt}
\ea
Using the identity $J_n(x) Y_{n+1}(x) - Y_n(x) J_{n+1}(x) = -\tfrac{2}{\pi x}$,
the expression for $|u(m,0)|^2$ can be simplified to
\ba
  |u(m,0)|^2 \eq \frac{4\mu^2}{\pi^2 m^2} \frac{N_m^2}{J_1^2(\tfrac{m}{\mu})}
    \nonumber \\
  \eq \frac{2\mu^3}{\pi^2 m^2} \left[
    \int_0^{\mu z_r} (1+x)
  \right. \nonumber \\
  && \hspace{-15mm} \times \! \left.
    \left\{
      J_1(\tfrac{m}{\mu}) Y_2[\tfrac{m(1+x)}{\mu}] -
      Y_1(\tfrac{m}{\mu}) J_2[\tfrac{m(1+x)}{\mu}]
    \right\}^2 dx
  \right]^{-1} . \nonumber \\ &&
  \label{eq:usqr}
\ea
In order to proceed further we must find the allowed mass eigenvalues by
solving (\ref{eq:eigenvalues}), at least approximately.

If we also include the radion, which has been ignored until now, the potential
changes to \cite{Callin2}
\ba
  V(r) \eq -\frac{G m_1 m_2}{r} \nonumber \\
  && \hspace{-5mm} \times \! \left\{
    1 + \frac{4}{3\mu} \frac{1 - e^{-2\mu y_r}}{1 + \frac{1}{3} e^{-2\mu y_r}}
      \sum_{m>0} |u(m,0)|^2 e^{-mr}
  \right\} , \hspace{6.5mm}
  \label{eq:V_eksakt_radion}
\ea
and the four-dimensional Planck mass is then
\be
  \MPl^2 = \frac{M^3}{\mu}
    \frac{1 - e^{-2\mu y_r}}{1 + \frac{1}{3} e^{-2\mu y_r}}
  \equiv \frac{1}{8\pi G} \, .
  \label{eq:4dPlanckmass_radion}
\ee
Comparing eq. (\ref{eq:V_eksakt_radion}) and (\ref{eq:4dPlanckmass_radion})
with (\ref{eq:V_eksakt}) and (\ref{eq:4dPlanckmass}), respectively, we see that
the whole effect of the radion is the extra factor \mbox{$(1 + \frac{1}{3}
e^{-2\mu y_r})$}.

\subsection{Finding the mass eigenvalues}

So far, the conformal distance $z_r$ to the second brane has been considered a
free parameter of arbitrary magnitude. However, the original idea of Randall
and Sundrum \cite{RandallSundrum1} was that the two-brane model should explain
the hierarchy between the electroweak scale $\MEW \sim 1 \; \mathrm{TeV}$ and
the Planck scale $\MPl \sim 10^{18} \; \mathrm{GeV}$, which is obtained by
setting
\be
  e^{\mu y_r} = 1 + \mu z_r \sim 10^{15} .
\ee
In the following, we will therefore use as a simplifying assumption that $\mu
z_r$ is a very large number. The general features of the results we will obtain
in the end, however, are independent of this assumption. In appendix
\ref{sec:smallz} the other extreme case $\mu z_r \ll 1$ is studied, showing
that the only thing that affects whether the two-brane model can be
distinguished from the one-brane model is the ratio $r/z_r$.

The function $f(x) \equiv Y_1(x) / J_1(x)$ is almost periodic, and approaches
$\tan(x - \tfrac{3\pi}{4})$ fast for large $x$. From (\ref{eq:eigenvalues}) we
therefore get $m_n z_r \approx n\pi$ as a first approximation. However, we can
find a much more accurate expression than this for the lowest eigenvalues in
the limit $\mu z_r \gg 1$. Since $\tfrac{m_n}{\mu} \approx \tfrac{n\pi}{\mu
z_r} \ll 1$ we expand $f(x)$ for small arguments:
\be
  f(\tfrac{m}{\mu}) =
    -\frac{4\mu^2}{\pi m^2} + {\cal O}(\ln\tfrac{m}{\mu}) \, .
\ee
The pole at $x=0$ must coincide with a pole at $x=c_n$, that is, the $n$-th
zero of the Bessel function, $J_1(c_n) = 0$. With the expansion $f(c_n +
\epsilon_n) = Y_1(c_n)/\epsilon_n J_1'(c_n) + {\cal O}(1)$ and demanding that
$f(\tfrac{m}{\mu}) = f[\tfrac{m}{\mu}(1+\mu z_r)] = f(c_n + \epsilon_n)$ we
then get
\be
  \epsilon_n = -\frac{\pi m_n^2 Y_1(c_n)}{4\mu^2 J'_1(c_n)} +
    {\cal O}(\tfrac{m_n^4}{\mu^4} \ln \tfrac{m_n}{\mu}) \, ,
\ee
and hence
\ba
  && \frac{m_n}{\mu} = \frac{c_n + \epsilon_n}{1 + \mu z_r} \nonumber \\
  && \hspace{-4mm} = \frac{c_n}{1+\mu z_r} -
    \frac{\pi Y_1(c_n) c_n^2}{4 J'_1(c_n) (1+\mu z_r)^3} + {\cal O}\!\left[
      \tfrac{c_n^4}{(1 + \mu z_r)^5} \ln\tfrac{c_n}{1 + \mu z_r}
    \right] \! . \nonumber \\
  && \label{eq:mass_small}
\ea
For large values of $n$ we use $J_1(c_n) \sim \cos(c_n - \tfrac{3\pi}{4}) = 0$,
which means that $c_n \simeq (n+\tfrac{1}{4})\pi$.

\subsection{Short distances, $\mu r \ll 1$}
\label{sec:shortdistances}

For very short distances we should expect to get the same result as for flat,
five-dimensional Minkowski space, i.e. we should expect the relative correction
$\Delta(r)$ to the Newtonian potential to be proportional to $1/r$, where $V(r)
= V_0(r)(1+\Delta)$, $V_0(r) = -G m_1 m_2 / r$. The full potential will then
behave like $V(r) \sim 1/r^2$ when $r \to 0$, since $\Delta \gg 1$ in this
limit.

From (\ref{eq:V_eksakt_radion}) we see that $r \to 0$ corresponds to letting $m
\to \infty$. More precisely, from the assumption $\mu r \ll 1$ we get $m \sim
r^{-1} \gg \mu$, allowing us to use the asymptotic expressions for the Bessel
functions in (\ref{eq:usqr}). The integral thus reduces to
\be
  I_m \simeq \frac{4\mu^2}{\pi^2 m^2}
    \int_0^{\mu z_r} \cos^2(\tfrac{mx}{\mu}) dx \simeq
  \frac{2\mu^3 z_r}{\pi^2 m^2},
\ee
which means that $|u(m,0)|^2 \simeq 1/z_r$. By also using $m_n \simeq n\pi/z_r$
we then get
\ba
  \Delta \eqsimeq \frac{4}{3\mu z_r}
    \frac{1 - e^{-2\mu y_r}}{1 + \frac{1}{3} e^{-2\mu y_r}}
    \sum_{n=1}^\infty e^{-n\pi r/z_r} \nonumber \\
  \eq \frac{4}{3\mu z_r} \frac{1-e^{-2\mu y_r}}{1+\frac{1}{3}e^{-2\mu y_r}}
    \frac{1}{e^{\pi r/z_r} - 1}.
\ea
Since $\mu r \ll 1$ implies $r \ll z_r$ we can expand the exponential in powers
of $r/z_r$, obtaining the final result
\be
  \Delta \simeq \frac{4}{3\pi\mu r}
    \frac{1 - e^{-2\mu y_r}}{1 + \frac{1}{3} e^{-2\mu y_r}} \, ,
    \hspace{8mm} \mu r \ll 1 \, .
  \label{eq:correction_smallr}
\ee
The only dependence on the brane separation $y_r$ lies in the factor
$(1-e^{-2\mu y_r})/(1+\frac{1}{3}e^{-2\mu y_r})$, which is exactly the factor
that enters into the five-dimensional Planck mass $M$ from
(\ref{eq:4dPlanckmass_radion}). To be more spesific, since $\Delta \gg 1$ the
complete potential is
\be
  V(r) \simeq -\frac{4G}{3\pi\mu}
    \frac{1-e^{-2\mu y_r}}{1+\frac{1}{3}e^{-2\mu y_r}} \frac{m_1 m_2}{r^2}
  = -\frac{m_1 m_2}{6\pi^2 M^3 r^2},
\ee
by using (\ref{eq:4dPlanckmass_radion}) and $G = (8\pi\MPl^2)^{-1}$. The
natural way of defining the gravitational constant $G_D$ in $D$ spacetime
dimensions is by requiring the gravitational force to be $F = G_D m_1 m_2 /
r^{D-2}$. Since the force is given by the derivative of the potential we thus
get $G_5 = (3\pi^2 M^3)^{-1}$, and hence $V(r) \simeq -G_5 m_1 m_2 / 2r^2$.

From (\ref{eq:correction_smallr}) we also notice, of course, that we reproduce
the one-brane result by taking the limit $y_r \to \infty$.

\subsection{Large distances, $\mu r \gg 1$}

The limit of very large distances, $r \to \infty$, corresponds to $m \to 0$ (or
more precisely $m \ll \mu$). Things will therefore be a little more
complicated, since we can't use the asymptotic expressions for the Bessel
functions in (\ref{eq:usqr}) anymore. Instead, the integral must be evaluated
more or less analytically as it stands. Changing to the variable $u =
\tfrac{m}{\mu}(1+x)$ we get
\ba
  I_m \eqsimeq
    \frac{\mu^2}{m^2} \int_{\frac{m}{\mu}}^{c_n} u \left[
      J_1(\tfrac{m}{\mu}) Y_2(u) - Y_1(\tfrac{m}{\mu}) J_2(u)
    \right]^2 du \nonumber \\
  \eqequiv \frac{\mu^2}{m^2} \! \left[
    J_1^2(\tfrac{m}{\mu}) I_1 -
    2 J_1(\tfrac{m}{\mu}) Y_1(\tfrac{m}{\mu}) I_2 +
    Y_1^2(\tfrac{m}{\mu}) I_3
  \right] \! . \hspace{6mm}
\ea
The two integrals $I_1 = \int u Y_2^2(u) du$ and $I_3 = \int u J_2^2(u) du$ are
easily expressed in terms of other Bessel functions, whereas the integral $I_2
= \int u Y_2(u) J_2(u) du$ is a little more complicated and involves the Meijer
$G$-function \cite{MeijerG}. However, all the different terms can be expanded
in powers of $m/\mu$, and they turn out to be of completely different orders of
magnitude. The dominant term can be shown to be the upper limit $u=c_n$ in
$I_3$, where $Y_1^2(\tfrac{m}{\mu}) \simeq \frac{4\mu^2}{\pi^2 m^2}$ and thus
\be
  I_m \simeq \frac{4\mu^4}{\pi^2 m^4} I_3(c_n) =
    \frac{2\mu^4 c_n^2 J_2^2(c_n)}{\pi^2 m^4},
\ee
where we have used $I_3(u) = \tfrac{1}{2}u^2 \left[ J_2^2(u) - J_1(u) J_3(u)
\right]$ together with $J_1(c_n) = 0$. Inserting this into (\ref{eq:usqr})
gives
\be
  |u(m,0)|^2 \simeq \frac{m^2}{\mu c_n^2 J_2^2(c_n)} \simeq
    \frac{\mu}{(1+\mu z_r)^2 J_2^2(c_n)}.
\ee

For very large distances, $r \gg z_r$, we get the dominant contribution from
the first eigenvalue $m_1$,
\ba
  \Delta \eqsimeq
    \frac{4}{3(1+\mu z_r)^2 J_2^2(c_1)}
    \frac{1-e^{-2\mu y_r}}{1+\frac{1}{3}e^{-2\mu y_r}} e^{-m_1 r}
    \nonumber \\
  \eqsimeq \frac{8.21953739}{(1+\mu z_r)^2} e^{-3.83170597 \, r/z_r} ,
    \hspace{12mm} r \gg z_r, \hspace{8mm}
  \label{eq:correction_verylarger}
\ea
where we have ignored the terms $e^{-2\mu y_r}$ (since this represents a higher
order correction) and inserted the value of $c_1$ in the last line. The
correction thus decreases exponentially for large distances, which is the
familiar result for a compact extra dimension. The correction is, however,
strongly suppressed compared to an ordinary compact dimension, because of the
factor $(1+\mu z_r)^{-2} \sim 10^{-30}$.

For intermediate distances we must sum over all the allowed eigenvalues. Using
the approximations $J_2^2(c_n) \simeq \tfrac{2}{\pi c_n}$ and $c_n \simeq
(n+\tfrac{1}{4})\pi$ we then get
\ba
  \Delta \eqsimeq \frac{2\pi^2}{3(1+\mu z_r)^2}
    \sum_{n=1}^\infty (n+\tfrac{1}{4}) e^{-(n+1/4)\pi r/z_r} \nonumber \\
  \eq \frac{\pi^2 e^{-\pi r / 4z_r}}{24(1+\mu z_r)^2}
    \frac{5-e^{-\pi r/z_r}}{\sinh^2(\tfrac{\pi r}{2z_r})} \, ,
    \hspace{8mm} \mu r \gg 1 \, . \hspace{5mm}
  \label{eq:correction_larger}
\ea
Since we have assumed $\mu z_r \gg 1$ this expression is also valid for $r \ll
z_r$, as long as we still have $\mu r \gg 1$. In this limit
(\ref{eq:correction_larger}) reduces to
\be
  \Delta \simeq \frac{2}{3\mu^2 r^2} \, ,
    \hspace{8mm} 1 \ll \mu r \ll \mu z_r \, ,
\ee
which is the same result as for a single brane \cite{Callin}, and in agreement
with \cite{Shtanov}.

\begin{figure}[!h]
  \begin{center}
    \includegraphics[width=75mm]{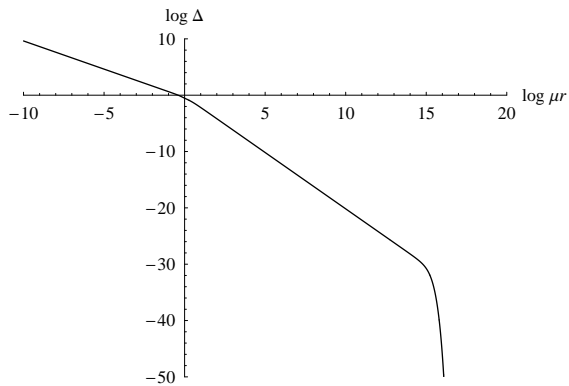}
  \end{center}
  \vspace{-5mm}
  \caption{The figure shows the relative correction $\Delta$ to the Newtonian
    potential from (\ref{eq:correction_smallr}) and
    (\ref{eq:correction_larger}) for short and large distances, respectively,
    where $V(r) = V_0(r) (1 + \Delta)$. The conformal distance $z_r$ to the
    second brane is chosen such that $\mu z_r = 10^{15}$. The figure clearly
    shows the transition from the one-brane result to that of a compact extra
    dimension around $\mu r \sim 10^{15}$, i.e. around $r \sim z_r$.}
  \label{fig:potential}
\end{figure}

In figure \ref{fig:potential} the two expressions (\ref{eq:correction_smallr})
and (\ref{eq:correction_larger}) for short and large distances are combined
into a single plot (using a simple interpolation for the transition at $\mu r
\sim 1$). As is evident from the figure, there are three different regions,
defined by the scale of the two parameters $\mu$ and $z_r$ in the theory: For
short distances ($\mu r < 1$) the potential is essentially five-dimensional,
just as expected, since spacetime can always be considered locally flat. For
intermediate distances ($\mu^{-1} < r < z_r$) we have the dominant term $\Delta
\simeq 2/3\mu^2 r^2$ just as in the one-brane model. Here the distance is large
enough to be affected by the warped geometry of the fifth dimension, but still
too short to notice the presence of the second brane. Finally, for large
distances ($r > z_r$) the fifth dimension is manifestly compact, with an
exponential cutoff in the correction to the potential.

This general shape of the correction $\Delta(r)$ will be the same for all
choices of the parameters $\mu$ and $z_r$. The only thing that will change when
varying $z_r$ (and keeping $\mu$ fixed) is the position where the exponential
cutoff occurs. Of course, if $\mu z_r < 1$ we will only see two different
regions, as in this case the branes are so close together that the bulk space
between them is essentially flat.

\section{Summary}

In this paper we have considered the Randall-Sundrum model with two branes. The
goal has been to study the effect on the Newtonian gravitational potential due
to the finite distance between the branes. As expected, when the distance $r$
in the potential is small compared to the brane separation $z_r$, the potential
is asymptotically the same as in the one-brane model where $z_r \to \infty$. In
a sense, a short-range measurement of gravity is unable to see what is going on
at a much larger scale.

It is not until the distance $r$ becomes comparable to $z_r$ that we can see
the effect of the finite distance between the branes, and the potential changes
from that of the one-brane model to that of a model with an ordinary compact
extra dimension, where the relative correction to the four-dimensional
potential decreases exponentially. The entire bulk-space dimension thus becomes
invisible to a long-range experiment in gravity.

Both of these examples illustrate a quite general feature of almost all of
physics -- that a measurement done at a particular scale can only reveal
information about the underlying physics at the approximately same scale. It
should be noted, however, that it is not the physical distance $y_r$ to the
second brane that sets this scale, but rather the conformal distance $z_r
\simeq e^{\mu y_r}/\mu$, since all graviton masses are suppressed by the same
factor $e^{\mu y_r}$.

\vspace{4mm}\textbf{Acknowledgement:} I wish to thank Finn Ravndal for useful
discussions. This work has been supported by grant no. NFR 153577/432 from the
Research Council of Norway.

\appendix

\section{The limit $\mu z_r \ll 1$}
\label{sec:smallz}

When the conformal distance to the second brane is small compared to the
curvature radius of the fifth dimension, we would expect the fifth dimension to
behave like an ordinary compact dimension at all scales. More precisely, we
should get the same potential $V(r)$ for all distances $r$ as with flat,
five-dimensional space, where one of the dimensions is compact with an
extension $L = 2z_r$.

When $\mu z_r \ll 1$ the conformal distance $z_r$ between the branes is equal
to the physical distance $y_r$, since $1+\mu z_r = e^{\mu y_r} \simeq 1+\mu
y_r$. In this limit even the first mass eigenvalue will be very large, i.e.
$m_1/\mu \gg 1$, and we can use the asymptotic expressions for all Bessel
functions in (\ref{eq:eigenvalues}). This yields
$\tan(\tfrac{m}{\mu}-\tfrac{3\pi}{4}) \simeq
\tan(\tfrac{m}{\mu}-\tfrac{3\pi}{4}+m z_r)$, with the solution
\be
  m_n \simeq \frac{n\pi}{z_r} \, , \hspace{6mm} \mu z_r \ll 1 \, .
  \label{eq:mass_smallz}
\ee
This result should be compared to (\ref{eq:mass_small}). For large values of
$n$ the two expressions are identical to the lowest order, i.e. when
(\ref{eq:mass_small}) is expanded in powers of $1/\mu z_r$. However, in
(\ref{eq:mass_small}) it is better to use $\tfrac{m_n}{\mu} \simeq
\tfrac{c_n}{1+\mu z_r}$ than $m_n \simeq \tfrac{c_n}{z_r}$ even when $\mu z_r
\gg 1$, because the next correction in the latter case would be $\sim 1/\mu^2
z_r^2$, whereas the next correction in (\ref{eq:mass_small}) is $\sim 1/(1+\mu
z_r)^3 \ll 1/\mu^2 z_r^2$.

The quantization (\ref{eq:mass_smallz}) is exactly what we get from a flat,
compact extra dimension, where the wavefunction is $u(z) \sim e^{imz}$ and the
periodic boundary condition gives $e^{imL} = 1$, with $L = 2z_r$.

With $m_n/\mu \gg 1$ we get $|u(m,0)|^2 \simeq 1/z_r$ just as in section
\ref{sec:shortdistances}, and the correction $\Delta$ to the Newtonian
potential is therefore
\ba
  \Delta \eqsimeq \frac{4}{3\mu z_r}
    \frac{1 - e^{-2\mu y_r}}{1 + \frac{1}{3} e^{-2\mu y_r}}
    \sum_{n=1}^\infty e^{-n\pi r/z_r}
  \simeq \frac{2}{e^{\pi r/z_r} - 1} \, , \hspace{7mm}
  \label{eq:pot_smallz}
\ea
where we have used $y_r \simeq z_r$. Again we get the same result as with a
flat, extra dimension (see \cite{Callin2}). For short and large distances,
respectively, (\ref{eq:pot_smallz}) is simplified to
\be
  \Delta \simeq \left\{ \begin{array}{lc}
    \ds \frac{2 z_r}{\pi r}, & r \ll z_r, \vs \\
    \ds 2 e^{-\pi r/z_r}, & r \gg z_r.
  \end{array} \right.
\ee
Although a little more camouflaged than in the limit $\mu z_r \gg 1$, the
result for $r \ll z_r$ is once again the same as for a single brane, since the
factor $z_r$ is absorbed into the five-dimensional Planck mass. Alternatively,
if we had kept the factor $(1-e^{-2\mu y_r})/(1+\frac{1}{3}e^{-2\mu y_r})$ in
(\ref{eq:pot_smallz}), we would get the exact same result as
(\ref{eq:correction_smallr}) for $r \ll z_r$.

In any case, the conclusion is that in the limit $r \ll z_r$ the two-brane
model is indistinguishable from the one-brane model, whereas in the limit $r
\gg z_r$ the extra dimension behaves like an ordinary, flat-space compact
dimension.


\begin{thebibliography}{10}
\bibitem{radion}
At this point, we are simply ignoring the radion field $\phi(x) \sim h_{44}$.
The component $h_{\mu 4}$, on the other hand, can be set to zero as a gauge
choice \cite{Smolyakov, Boos}. The radion will later be inserted back into the
final result, using \cite{Callin2}.
\end{thebibliography}
\end{document}